\documentclass{WileyMSP-template}
\usepackage{amsmath}
\usepackage{xcolor}
\def\_#1{{\bf #1\mit}}

\begin{document}

\pagestyle{fancy}
\rhead{\includegraphics[width=2.5cm]{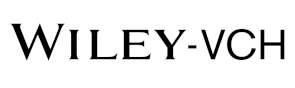}}


\title{Stacked Time-Varying Metasurfaces}

\maketitle


\author{Mostafa Movahediqomi}
\author{Sergei Tretyakov}
\author{Viktar Asadchy*}
\author{Xuchen Wang}


\dedication{}

\begin{affiliations}
Mostafa Movahediqomi, Sergei Tretyakov, Viktar Asadchy\\
Department of Electronics and Nanoengineering\\
Aalto University\\
15500, FI-00076 Aalto, Finland.\\
Email Address: viktar.asadchy@aalto.fi

Xuchen Wang\\
Qingdao Innovation and Development Base\\
Harbin Engineering University\\
266400, Qingdao, China.\\

\end{affiliations}



\begin{abstract}

Spatiotemporal metasurfaces offer unique opportunities for wave manipulation, however, their practical realization is often constrained by the requirement for in-plane spatial modulation, which necessitates a large number of time-varying elements. In this work, we introduce an alternative architecture based on a cascade of spatially uniform metasurfaces subjected to periodic temporal modulation. Although all metasurfaces share the same modulation frequency, their individual modulation functions are independently engineered to achieve a desired complex electromagnetic response. 
We develop a general theoretical framework for the design and optimization of such stacked metasurface systems, composed of dense arrays of cylindrical meta-atoms with time-varying plasma and/or collision frequencies. The effectiveness of the approach is demonstrated through the optimization of metasurface designs that enable magnet-free isolation at the fundamental frequency and a temporal analogue of circulators. Furthermore, we show that a metasurface stack can be implemented using only a few time-modulated elements embedded within a parallel-plate waveguide, opening new avenues for extremely compact, versatile, and scalable spatiotemporal platforms for next-generation photonic and microwave systems.

\end{abstract}


\section{Introduction}

In recent years, the pursuit of artificial materials with tailored electromagnetic responses has led to the rapid development of metamaterials and, more recently, metasurfaces~\cite{achouri2021electromagnetic}. Traditionally, such structures were designed exclusively through spatial engineering, by modifying the geometry, size, or arrangement of the constituent scatterers. Yet, this approach is inherently restricted. For time-invariant and linear scatterers, the scattered frequency and total energy are conserved, preventing access to many  desirable functionalities. These limitations have spurred growing interest in exploiting the temporal dimension of material parameters, which provides a powerful route to fundamentally new scattering behaviors~\cite{caloz2019spacetime, mostafa2024temporal,asgari2024theory}.

Time-varying platforms have already revealed the potential to unlock unconventional wave phenomena, such as unbounded field accumulation~\cite{mirmoosa2019time} and temporal analogues of Wood’s anomalies~\cite{galiffi2020wood}. In this context, spatiotemporal metasurfaces have emerged as a particularly promising architecture, enabling frequency conversion~\cite{shaltout2015time, buddhiraju2021arbitrary}, isolation~\cite{wang2020theory,cardin2020surface}, circulation~\cite{shi2017optical,wang2020theory}, power combination~\cite{wang2021space}, and nonreciprocal phase manipulation~\cite{wang2020theory}. The central principle is the induction of a synthetic motion along the metasurface plane, which effectively breaks time-reversal symmetry~\cite{mazor2019one}. However, realizing such metasurfaces remains technically demanding. Approaches based on continuous space-time gradients require carefully engineered impedance profiles~\cite{hadad2015space}, while discrete space-time metagratings rely on dense arrays of synchronized meta-atoms~\cite{hadad2024space}—both posing major implementation barriers due to the in-plane spatial modulation requiring non-local design and many time-varying elements.

In this work, we introduce a different strategy that mitigates these challenges by shifting the spatial modulation into the out-of-plane direction. Our system consists of a stack of metasurfaces that are uniform across their planes but undergo periodic modulation in time. This longitudinal arrangement enhances the design freedom compared to conventional in-plane spatiotemporal metasurfaces, while providing easier practical realization. Our theoretical modeling is performed in terms of bulk parameters of the meta-atoms, such as plasma and collision frequencies, with explicit consideration of frequency dispersion—an aspect particularly advantageous for optical realizations. Conceptually, the proposed stacked structures resemble traveling-wave modulation in transmission-line and optical systems~\cite{cullen1958travelling,Yu2009} or tandem-type modulation~\cite{fang2012photonic}, yet they are more general and substantially more compact. Furthermore, these metasurface cascades represent a generalization of recently proposed single-layer arrays of time-modulated spherical meta-atoms~\cite{garg2022modeling, garg2025inverse}.
We showcase their versatility by designing stacked metasurfaces capable of magnet-free isolation at the fundamental frequency and a temporal analogue of circulators. Remarkably, both designed metasurface stacks have electrical thickness below the operational wavelength. Furthermore, we demonstrate that such architectures can be straightforwardly implemented in parallel-plate waveguides with only a few time-modulated elements.

\section{Theoretical framework}
\label{sec:Theory}


\begin{figure}
\centering
  \includegraphics[width=0.9\linewidth]{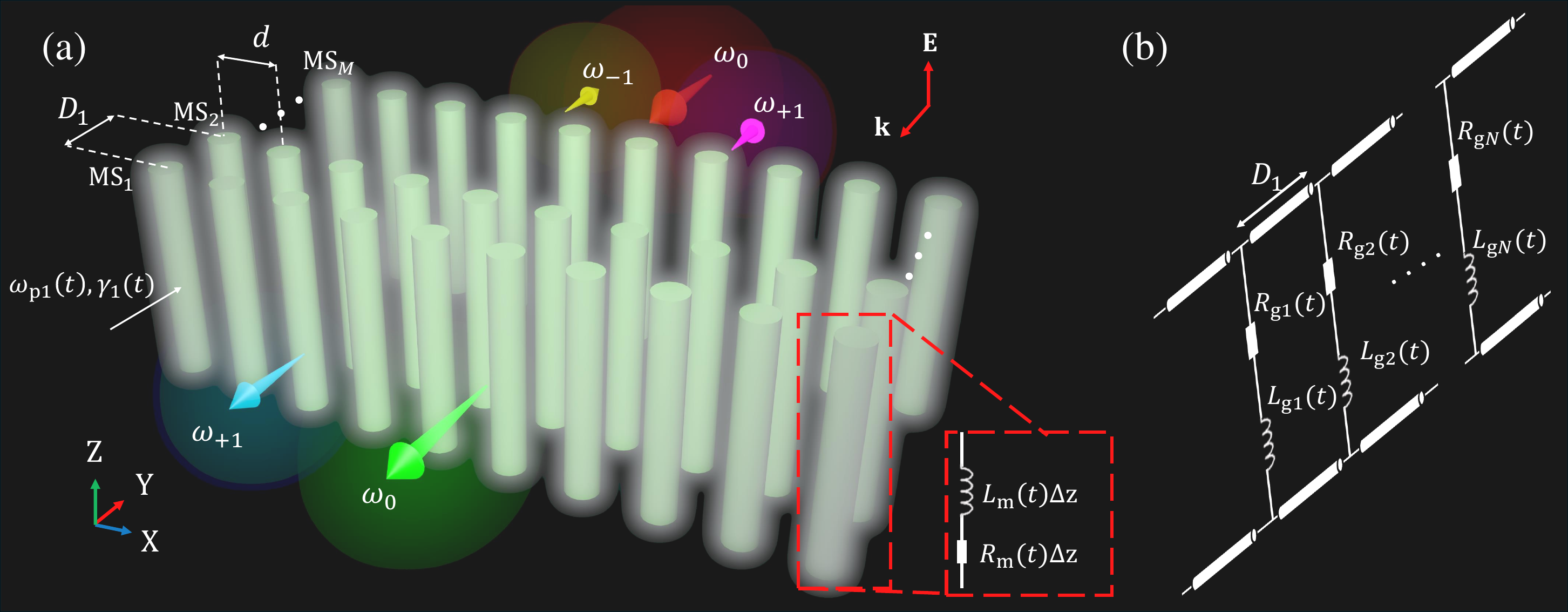}
  \caption{(a) Geometry of the considered cascade of time-varying metasurfaces (denoted for brevity as ${\rm MS}_1$, ${\rm MS}_2$, ... ${\rm MS}_M$). Each metasurface consists of a dense array of thin long cylinders which makes it effectively uniform in the $xz$-plane with material properites (plasma and/or collision frequencies) periodically varying in time. The modulation functions of different metasurfaces are designed independently. The red dashed inset denotes that each cylindrical meta-atom can be modeled by a series connection of time-varying inductance and resistance per unit length.
  (b) Equivalent transmission-line model of the metasurface stack shown in (a) consisting of time-varying shunt resistances and inductances separated by sections of the transmission lines modeling free-space gaps between the metasurfaces.}
  \label{fig:conceptual_figure}
\end{figure}

Let us consider the configuration of stacked time-varying metasurfaces shown in Fig.~\ref{fig:conceptual_figure}(a), illuminated by a plane wave along the normal direction. Each metasurface comprises a dense array of long cylinders with material properties varying periodically in time. The cylinders have extent along the $z$-direction much larger than the operational wavelength and, therefore, are modeled as infinitely long. The incident-wave electric field is polarized along the cylinders. The cylinder radii $r_0$ and the spacing between the neighboring cylinders $d$ are both assumed to be sub-wavelength to ensure that each metasurface is effectively uniform in the $xz$-plane. The separation distance between the $i$-th and $(i+1)$-th metasurfaces is denoted as $D_i$ and can be arbitrary but larger than $d$ to ensure that there is no near-field coupling between the cylinders in neighboring metasurfaces~\cite{tretyakov2003analytical}. 
We model the material of each cylinder in the $i$-th metasurface with Drude dispersion with periodically time-varying plasma frequency  $\omega_{{\rm p},i}^2(t)=\frac{N_i(t)e^2}{m \epsilon_0}$ ($N_i(t)$ is the time-dependent electron density, $m$ and $e$ are the effective electron mass and electric charge) and periodically time-varying electron collision frequency $\gamma_i(t)$. It is important to note that since such cylinder arrays can be considered as impedance sheets, other physical realizations are equally possible.

\subsection{Modeling a single time-varying cylindrical meta-atom}
We begin by analyzing a single cylinder with time-varying material properties. Here, for brevity, we omit the use of the subscript $i$ denoting which metasurface in the stack the considered meta-atom belongs to. 
The dynamics of its polarization density $\_P$ is governed by
\begin{equation}
\frac{{\rm d}^2 \_P (t)}{{\rm d}t^2} + \gamma (t) \frac{{\rm d}\_P(t)}{{\rm d}t} = \omega_{\rm p}^2(t) \, \varepsilon_0 \_E(t),
\label{eq:time_Drude}
\end{equation}
where  $\_E(t)$ is the total electric field in the cylindrical meta-atom. 
Since the current density inside the meta-atom is related to its bulk electric polarization as $\_J_{\rm m}(t) = \frac{{\rm d}\_P(t)}{{\rm d}t}$, Eq.~(\ref{eq:time_Drude}) can be rewritten as
\begin{equation}
    \frac{{\rm d} \_J_{\rm m}(t)}{{\rm d} t}+ \gamma(t) \_J_{\rm m}(t)=\omega_{\rm p}^2(t) \epsilon_0 \_E(t). \label{eq: differential}
\end{equation}
Assuming that the current density is approximately uniform over the cylinder cross section, the total current in the meta-atom is the product of the current density and the cross-sectional area $S=\pi r_0^2$, that is, $I_{\rm m}(t)=J_{\rm m}(t) S$.
Substituting  this into Eq.~(\ref{eq: differential}) and taking the projection of the equation to the $z$-axis yields:
\begin{equation}
        \frac{1}{\omega_{\rm p}^2(t) \epsilon_0 S} \frac{{\rm d} I_{\rm m}(t)}{{\rm d} t}+ \frac{\gamma(t)}{\omega_{\rm p}^2(t) \epsilon_0 S} I_{\rm m}(t)=E(t). \label{eq: differential2}
\end{equation}
We can notice that Eq.~(\ref{eq: differential2}) can be interpreted as an equation describing voltage and current in a series connection of time-varying resistance $R_{\rm m}(t)$ and time-varying inductance $L_{\rm m}(t)$: 
\begin{equation}
    \frac{\rm d}{{\rm d}t}\left[L_{\rm m}(t) I_{\rm m}(t)\right] + R_{\rm m}(t) I_{\rm m}(t) = v(t)/\Delta z,
\end{equation}
where $\Delta z$ is the differential length of the cylinder along the $z$-direction.
Therefore, a single time-varying cylindrical meta-atom can be modeled by a series connection of time-varying inductance and resistance given by 
\begin{equation}
    L_{\rm m}(t)=\frac{1}{\omega_{\rm p}^2(t) \epsilon_0 S}, \quad R_{\rm m}(t)=\frac{\gamma(t)}{\omega_{\rm p}^2(t) \epsilon_0 S} - \frac{{\rm d} }{{\rm d} t} \left[\frac{1}{\omega_{\rm p}^2(t) \epsilon_0 S}\right].
    \label{eq:cell_component}
\end{equation}
Note that the physical meaning of $L_{\rm m}$ and $R_{\rm m}$ are the resistance and inductance per unit length of the cylindrical meta-atom, with units of [H/m] and [$\Omega$/m], respectively. The equivalent circuit of a differential length $\Delta z$ of the meta-atom is shown in Fig.~\ref{fig:conceptual_figure}(a). 

\subsection{Modeling an array of time-varying cylindrical meta-atoms}
The above relations were derived for a single meta-atom. We now extend the analysis to a dense array of such meta-atoms with periodicity $d$. 
If the meta-atoms were time-invariant, the grid impedance of such an array would be given by~\cite[Eq.~(4.39)]{tretyakov2003analytical}
\begin{equation}
 Z_{\rm g}=  Z_{\rm m}d + j \dfrac{\eta_0}{2} \alpha_{\rm ABC},  \label{eq: grid impedance}  
\end{equation}
where $Z_{\rm m}= R_{\rm m}+j \omega  L_{\rm m}$ is the complex impedance per unit length of a single meta-atom defined in the frequency domain, $j$ is the imaginary unit (we adopt $e^{j\omega t}$ notation for harmonic oscillations), and $\eta_0=\sqrt{\mu_0/\varepsilon_0}$ is the wave impedance of the background material (here assumed to be free space). Parameter $\alpha_{\rm ABC} = \dfrac{k_0 d}{\pi} \ln \left( \dfrac{d}{2\pi r_0} \right)$ is the grid parameter established by Kontorovich et al.~\cite{kontorovich1962reflection} in the framework of averaged boundary conditions (here $k_0$ is the wavenumber of the background material). It is related to the frequency and geometry of the array, but independent of the material properties. 
The additional inductive term in Eq.~(\ref{eq: grid impedance}) arises purely from the geometry. 

For the array of time-varying meta-atoms,  this geometric contribution must also be included. Therefore, the equivalent grid resistance and inductance of the metasurface is given by
\begin{subequations}
    \begin{equation}
    R_{\mathrm{g}}(t) = R_{\rm m}(t)d,
\label{eq:resistance_definition_grid}
\end{equation}
\begin{equation}
    L_{\mathrm{g}}(t) =\left[ L_{\rm m}(t)+\frac{\eta_0}{2}\frac{\sqrt{\epsilon_0 \mu_0}}{\pi}\ln\left( \frac{d}{2\pi r_0}\right) \right ]d.
\end{equation} \label{eq: grid impedance_timevarying}
\end{subequations}
Note that here  $R_{\mathrm{g}}(t)$  and $L_{\mathrm{g}}(t)$ have units of [$\Omega$]  and [H]. Thus, the stacked metasurfaces can be modeled by a cascade of shunt resistances and inductances in an equivalent transmission-line model shown in Fig.~\ref{fig:conceptual_figure}(b). The spacing between neighboring metasurfaces is modeled by a finite-length section of a transmission line with characteristic impedance equal to the wave impedance of the background medium. In Supplementary Material, we compare the transmission-line model developed here with the full-wave simulations of the metasurface.


\subsection{Mode-matching method}

So far, we have considered a generic, not necessarily periodic, temporal variation of material properties within the meta-atoms. We now focus on the special case of \textit{periodic} temporal modulations of material parameters. Specifically, we aim to analyze the reflection and transmission of different frequency harmonics from a single metasurface subject to a \textit{time-periodic} modulation of the plasma frequency $\omega_{\rm p}$ and the collision frequency $\gamma$ of its constituent meta-atoms. Given the temporal profiles of $\omega_{\rm p}$ and $\gamma$, we derive the corresponding time-varying grid resistance $R_{\rm g}$ and inductance $L_{\rm g}$. Employing the mode-matching method~\cite{wang2020nonreciprocity}, we subsequently determine the ABCD (transfer) matrix of the time-varying metasurface. This matrix formalism can be readily extended to compute the reflection and transmission characteristics of arbitrarily complex stacks composed of multiple metasurfaces.

The material parameters $\omega_{{\rm p}}$ and $\gamma$  of different metasurfaces and their temporal modulation profiles can be chosen independently. Nevertheless, in our analysis, they all share the same fundamental modulation frequency $\omega_{\rm m}$, resulting in the fact that the entire stack of metasurfaces has a temporal periodicity $T=2\pi/\omega_{\rm m}$. Therefore, we expand the material parameters of the $i$-th metasurface into the Fourier series:
\begin{subequations}
\begin{align}
\omega_{{\rm p},i}(t) &= \sum_{q=-\infty}^{+\infty} w_{q,i} e^{j q \omega_{\rm m} t} \label{eq:fourier_series_a} \\
\gamma_i(t) &= \sum_{q=-\infty}^{+\infty} g_{q,i} e^{j q \omega_{\rm m} t}, \label{eq:fourier_series_b}
\end{align}
\end{subequations}
where $w_{q,i}$ and $g_{q,i}$ determine the harmonic coefficients for the $i$-th metasurface. 




Since $R_{{\rm g},i}(t)$ and $L_{{\rm g},i}(t)$ are the functions of 
$\omega_{{\rm p},i}(t)$ and $\gamma_i(t)$, they are also periodic with the same period $T$ and can be written as
\begin{subequations}
\begin{align}
R_{{\rm g},i}(t) = \sum_{q=-\infty}^{+\infty} r_{q,i} e^{j q \omega_{\rm m} t}
\label{eq:fourier_series_aaaa} \\
L_{{\rm g},i}(t) = \sum_{q=-\infty}^{+\infty} l_{q,i} e^{j q \omega_{\rm m} t}.
\label{eq:fourier_series_bbb}
\end{align}
\end{subequations}

The harmonic coefficients $r_{q,i}$ and $l_{q,i}$ can be expressed in terms of $w_{q,i}$ and $g_{q,i}$ using Eq.~(\ref{eq:cell_component}). 
Below, for brevity, we omit the metasurface index ``$i$'', with all quantities henceforth referring to the same chosen metasurface.

The modulation induces an infinite number of harmonics with the
frequencies $\omega_n=\omega_0+n \omega_{\rm m}$, where $\omega_0$ is the frequency of the incident on the metasurface plane wave. Therefore, the current going through and the voltage across the equivalent resistor and inductor can be written as 
\begin{equation}
\begin{array}{l}
\displaystyle  I_{\rm L}(t) = I_{\rm R}(t)  = \sum_{n=-\infty}^{+\infty} i_n^{\rm L} e^{j \omega_n t}, \vspace{2mm} \\
\displaystyle V_{\rm L,R}(t) = \sum_{n=-\infty}^{+\infty} v_n^{\rm L,R} e^{j \omega_n t}.
\end{array}
\label{eq:current_voltage_fourier}
\end{equation}

In conventional static  circuits, the voltage and current are scalar values. Moreover, the
impedance of lumped components, which relates the scalar current and voltage, is a scalar.
However, in this multi-mode system, the voltage and current can be written in the form
of vectors, i.e., 
\[
\_v^{\rm L,R} = [\cdots, v^{\rm L,R}_{-1}, v^{\rm L,R}_{0}, v^{\rm L,R}_{+1}, \cdots]^T
\quad \text{and} \quad
\_i^{\rm L} = [\cdots, i^{\rm L}_{-1}, i^{\rm L}_{0}, i^{\rm L}_{+1}, \cdots]^T ,
\]
where each element in the vector denotes the complex amplitude of each harmonic. The
impedance or admittance of the modulated components should be a matrix which relates
the current and voltage vectors to one another. In order to calculate the scattering
harmonics, we should know the matrix representations of the resistor and inductor under modulation.  

The current-voltage relation of the time-varying inductance in the time domain is given by 
\begin{equation}
I_{\rm L}(t) L_{\rm g}(t) = \int^{t} V_{\rm L}(t) \, {\rm d}t.
\label{eq:inductor_integral}
\end{equation}

By substituting~Eqs.~\eqref{eq:current_voltage_fourier}, and~\eqref{eq:fourier_series_bbb} in~Eq.~\eqref{eq:inductor_integral}, we can obtain the corresponding equation in  matrix form as $\overline{\overline{Z}}_{\rm L} \cdot \_i^{\rm L} = \_v^{\rm L}$ or explicitly as~\cite{wang2020nonreciprocity}
\begin{equation}
\left(
\begin{array}{cccc}
j l_0 \omega_{-N} & j l_{-1} \omega_{-N} & \cdots & j l_{-2N} \omega_{-N} \\
j l_1 \omega_{1-N} & j l_0 \omega_{1-N} & \cdots & \vdots \\
\vdots & \vdots & \ddots & \vdots \\
j l_{2N} \omega_{N} & j l_{2N-1} \omega_{N} & \cdots & j l_0 \omega_{N}
\end{array}
\right) \cdot
\left(
\begin{array}{c}
i^{\rm L}_{-N} \\
i^{\rm L}_{1-N} \\
\vdots \\
i^{\rm L}_{N}
\end{array}
\right)
=
\left(
\begin{array}{c}
v^{\rm L}_{-N} \\
v^{\rm L}_{1-N} \\
\vdots \\
v^{\rm L}_{N}
\end{array}
\right),
\label{eq:matrix_relation}
\end{equation}
when truncating the Floquet harmonics  $n$ beyond the range from $-N$ to $N$.


For the resistance, the relation between the current and voltage of harmonics has a simple form:
\begin{equation}
V_{\rm R}(t) = R_{\rm g}(t) \, I_{\rm L}(t).
\label{eq:ohms_law_time_varying_indexed}
\end{equation}
Following the same procedure, we obtain the following matrix equation: 
\begin{equation}
\left(
\begin{array}{cccc}
r_0 & r_{-1} & \cdots & r_{-2N} \\
r_1 & r_0 & \cdots & \vdots \\
\vdots & \vdots & \ddots & \vdots \\
r_{2N} & r_{2N-1} & \cdots & r_0
\end{array}
\right) \cdot
\left(
\begin{array}{c}
i^{\rm L}_{-N} \\
i^{\rm L}_{1-N} \\
\vdots \\
i^{\rm L}_{N}
\end{array}
\right)
=
\left(
\begin{array}{c}
v^{\rm R}_{-N} \\
v^{\rm R}_{1-N} \\
\vdots \\
v^{\rm R}_{N}
\end{array}
\right),
\label{eq:resistance_matrix_relation}
\end{equation}
which can be written in a short form as  
$\overline{\overline{Z}}_{\rm R} \cdot \_i^{\rm L} = \_v^{\rm R}$. 

The total impedance matrix of the metasurface is a summation of the inductance and resistance matrices, that is,  $\overline{\overline{Z}}_{\mathrm{g}} = \overline{\overline{Z}}_{\rm L} + \overline{\overline{Z}}_{\rm R}$. 

\subsection{Transfer-matrix method}
For the general case when multiple metasurfaces are arranged in a cascade, it is convenient to define the transfer (ABCD) matrix that relates the currents and voltages at the input and output of a given metasurface as 
\[
\begin{pmatrix}
\_v^{\text{inp}}_{\text{tot}} \\
\_i^{\text{inp}}_{\text{tot}}
\end{pmatrix}
=
\begin{pmatrix}
\overline{\overline{A}} & \overline{\overline{B}}\\
\overline{\overline{C}} & \overline{\overline{D}}
\end{pmatrix} \cdot
\begin{pmatrix}
\_v^{\text{out}}_{\text{tot}} \\
\_i^{\text{out}}_{\text{tot}}
\end{pmatrix}
= \overline{\overline{T}}\cdot
\begin{pmatrix}
\_v^{\text{out}}_{\text{tot}} \\
\_i^{\text{out}}_{\text{tot}}
\end{pmatrix}
.
\]

The transfer matrix for the $i$-th metasurface can be described as~\cite[Sec.~4.4]{pozar2021microwave}:
\begin{equation}
\overline{\overline{T}}_{\mathrm{m},i} =
\left(
\begin{array}{cc}
\overline{\overline{I}} & 0 \\
\overline{\overline{Y}}_{\mathrm{g},i} & \overline{\overline{I}}
\end{array}
\right),
\label{eq:T_matrix_definition}
\end{equation}

where $\overline{\overline{I}}$ is the $(2N +1) \times (2N +1)$ unity matrix and $\overline{\overline{Y}}_{\mathrm{g},i}=\overline{\overline{Z}}_{\mathrm{g},i}^{-1}$.

The contribution of the air gaps between the neighboring metasurfaces should also be taken into account in the calculation. These gaps are equivalent to transmission-line sections with lengths $D_i$ and  wavenumbers for each Floquet harmonic $k_{n}=\omega_n/c$, where $c$ is the speed of light in free space. The transfer matrix of the $i$-th air gap is given by~\cite[Sec.~4.4]{pozar2021microwave}

\begin{equation}
\overline{\overline{T}}_{{\rm a},i} =
\left(
\begin{array}{cc}
\overline{\overline{I}} {\rm cos} \, k_n D_i  & \overline{\overline{I}} j \eta_0 \, {\rm sin} \, k_n D_i \\
\overline{\overline{I}} j \eta_0^{-1} \, {\rm sin} \, k_n D_i & \overline{\overline{I}} {\rm cos} \, k_n D_i
\end{array}
\right).
\label{eq:T_d_matrix}
\end{equation}

Eventually, the transfer matrix of the entire metasurface stack for forward and backward wave illuminations can be calculated by a multiplication of the transfer matrices:
\begin{equation}
\overline{\overline{T}}_{\rm F} = \overline{\overline{T}}_{\mathrm{m},1} \cdot \overline{\overline{T}}_{\rm a,1} \cdot 
\overline{\overline{T}}_{\mathrm{m},2} \cdot
\overline{\overline{T}}_{\rm a,2} \cdot \ldots \cdot
\overline{\overline{T}}_{{\rm a},M-1} \cdot
\overline{\overline{T}}_{\mathrm{m},M},
\label{eq:M_forward}
\end{equation}

\begin{equation}
\overline{\overline{T}}_{\rm B} = \overline{\overline{T}}_{\mathrm{m},M} \cdot \overline{\overline{T}}_{\rm a,M-1} \cdot 
\overline{\overline{T}}_{\mathrm{m},M-1} \cdot
\overline{\overline{T}}_{\rm a,M-2} \cdot \ldots \cdot
\overline{\overline{T}}_{{\rm a},1} \cdot
\overline{\overline{T}}_{\mathrm{m},1},
\label{eq:M_backward}
\end{equation}
where $M$ is the number of stacked metasurfaces.


Finding the transmission and reflection coefficients of the metasurface cascade, we should remember that the transfer matrix relates the total voltages and currents at the input and output. In other words, at the input, they include both incident and reflected waves, while at the output, only the transmitted waves:
\begin{equation}
\left(
\begin{array}{c}
\_v_{\rm i} + \_v_{\rm r} \\
 \eta_0^{-1}  (\_v_{\rm i} - \_v_{\rm r})
\end{array}
\right)
=
\overline{\overline{T}}_{\rm F} \cdot
\left(
\begin{array}{c}
\_v_{\rm t} \\
\eta_0^{-1} \_v_{\rm t}
\end{array}
\right)
=
\begin{pmatrix}
\overline{\overline{A}}_{\rm F} & \overline{\overline{B}}_{\rm F}\\
\overline{\overline{C}}_{\rm F} & \overline{\overline{D}}_{\rm F}
\end{pmatrix} \cdot 
\left(
\begin{array}{c}
\_v_{\rm t} \\
\eta_0^{-1} \_v_{\rm t}
\end{array}
\right).
\label{eq:interface_matching}
\end{equation}
The transmission and reflection coefficients for forward illumination scenario, defined as $\_v_{\rm t} = \overline{\overline{\tau}}_{\rm F} \cdot \_v_{\rm i}$ and $\_v_{\rm r} = \overline{\overline{\Gamma}}_{\rm F} \cdot \_v_{\rm i}$, can be derived as~\cite[Sec.~4.4]{pozar2021microwave}
\begin{equation}
\overline{\overline{\tau}}_{\rm F} =
2 \left( 
\overline{\overline{A}}_{\rm F} + \eta_0^{-1} \overline{\overline{B}}_{\rm F}  
+ \eta_0 \overline{\overline{C}}_{\rm F} 
+   \overline{\overline{D}}_{\rm F}  
\right)^{-1}
\label{eq:T_expression}
\end{equation}

\begin{equation}
\overline{\overline{\Gamma}}_{\rm F} =
\left( \overline{\overline{A}}_{\rm F} + \eta_0^{-1}\overline{\overline{B}}_{\rm F} \right)
\overline{\overline{\tau}}_{\rm F} - \overline{\overline{I}}.
\label{eq:R_expression}
\end{equation}

Likewise, one can obtain similar relations for the transmission and reflection coefficients for the backward illumination scenario $\overline{\overline{\tau}}_{\rm B}$ and $\overline{\overline{\Gamma}}_{\rm B}$. 

In summary, the developed theoretical formalism allows us to calculate complex amplitudes of all reflected and transmitted Floquet harmonics as described by Eqs.~(\ref{eq:T_expression}) and (\ref{eq:R_expression}) for a known material modulation given by Eqs.~(\ref{eq:fourier_series_a}) and (\ref{eq:fourier_series_b}). 
In the next section, we apply this formalism to an optimization algorithm to determine optimal parameters of temporal modulation of each metasurface for obtaining desired reflection and transmission responses of the metasurface cascade.

\section{Demonstrators}

Although the proposed theoretical framework is general and applicable to metasurface stacks with an arbitrary number of layers $M$ and arbitrary inter-metasurface distances $D_i$, for simplicity, we focus in this section on demonstrator designs, each consisting of two metasurfaces. As a form of temporal modulation in both metasurfaces, we assume a sinusoidal variation in the plasma and collision frequencies:
\begin{subequations}
\begin{align}
\omega_{{\rm p},1}(t) &= \bar{w}_1 \left[1 + \delta_1 \cos \omega_{\rm m} t \right] \label{eq:time_variant_a} \\
\gamma_{1}(t) &= \bar{\gamma}_1 \left[1 + \delta_1 \cos \omega_{\rm m} t \right] \label{eq:time_variant_b} \\
\omega_{{\rm p},2}(t) &= \bar{w}_2 \left[1 + \delta_2 \cos \left(\omega_{\rm m} t + \phi\right)\right] \label{eq:time_variant_c} \\
\gamma_{2}(t) &= \bar{\gamma}_2 \left[1 + \delta_2 \cos \left(\omega_{\rm m} t + \phi \right)\right] \label{eq:time_variant_d}
\end{align}
\end{subequations}
The coefficients defined in these equations can be uniquely converted to those in Eqs.~(\ref{eq:fourier_series_a}) and (\ref{eq:fourier_series_b}).
Without loss of generality, here we assume that the relative modulation strengths $\delta_{1,2}$ and modulation phases of both plasma and collision frequencies are the same for a given metasurface. We also assume a phase difference $\phi$ between the material properties of the two metasurfaces. 
Coefficients shown with the bars on top denote the time-average (mean) values of the corresponding material parameters.  

\subsection{Isolation at the fundamental frequency}

The quest of designing isolators using time-varying systems has been an active field of recent research~\cite{asadchy2020tutorial,williamson2020integrated,Kord2020}. Efficient isolation (nonreciprocal transmission) has been typically achieved in bulk systems where the electrical thickness is much larger than the operational wavelength~\cite{Yu2009,chamanara2017optical}. The challenge
of obtaining isolation in compact time-varying systems is difficult to overcome, despite relying on simultaneous frequency  conversion~\cite{zhang2019breaking, guo2019nonreciprocal} or the use of bianisotropy~\cite{wang2020nonreciprocity}. The former feature is undesired, as many applications require isolation at the fundamental (the same as the incident) frequency. The latter approach, although known, has not had practical realization to date.
Here, we realize isolation at the fundamental frequency in the stacked metasurface geometry depicted in Fig.~\ref{fig:conceptual_figure}(a). As we show below, only two metasurfaces inside the stack are sufficient to obtain isolation with high efficiency and without frequency conversions. This two-metasurface cascade has an effective time-varying bianisotropic response~\cite{mostafa2024temporal}.

For the optimization of material parameters, the \textit{fmincon} optimizer in MATLAB is selected due to its ability to handle nonlinear constraints and its compatibility with the ``MultiStart" setting, which enhances convergence speed and quality. The optimized parameters include those listed in Eqs.~(\ref{eq:time_variant_a})--(\ref{eq:time_variant_d}), that is, $\bar{w}_{1,2}$, $\bar{\gamma}_{1,2}$, $\delta_{1,2}$, $\phi$, and $\omega_{\rm m}$, as well as the distance between the layers $D_1 \gg d$. 
Without loss of generality, we fix $d=\lambda_0/20$ and the radius of the cylinders $r_0 = \lambda_0/150$, where $\lambda_0$ is the wavelength in free space at the fundamental (incident) frequency $\omega_0$. 

The optimization objective is to reach unit and zero amplitudes of the transmitted waves for the forward and backward illuminations, respectively, at the fundamental frequency $\omega_0$. The incident wave is represented by a single frequency $\omega_0$ with an electric field amplitude of 1~V/m in both illumination scenarios.
As mentioned before, the reflected and transmitted propagating waves in the considered system shown in Fig.~\ref{fig:conceptual_figure}(a) can be interpreted as the corresponding voltages in the equivalent transmission-line model, as depicted in Fig.~\ref{fig:conceptual_figure}(b). In other words, we can relate harmonics of the reflected and transmitted  electric fields to the incident fields using the same scattering coefficients as calculated from Eqs.~(\ref{eq:R_expression}) and (\ref{eq:T_expression}), i.e.,
$\_E_{\rm t}^{\rm F} = \overline{\overline{\tau}}^{\rm F} \cdot \_E_{\rm i}^{\rm F}$, and $\_E_{\rm t}^{\rm B} = \overline{\overline{\tau}}^{\rm B} \cdot \_E_{\rm i}^{\rm B}$. 
Thus, we define the cost function in the optimization as
\begin{equation}
\Delta_{\rm is} = \left| \left| E_{\rm t,0}^{\rm F} \right| - 1 \right| + \left| \left| E_{\rm t,0}^{\rm B}   \right| - 0 \right|.
\label{eq:error_metric}
\end{equation}
In this way, the optimizer tries to reach unit forward transmission at the fundamental frequency $\omega_0$, while minimizing the backward transmission for the same frequency harmonic. We truncate the range of Floquet harmonics to $n \in[-N; N]$ with $N=10$. The optimized material parameters are listed in Supplementary Material.

\begin{figure}
  \includegraphics[width=0.95\linewidth]{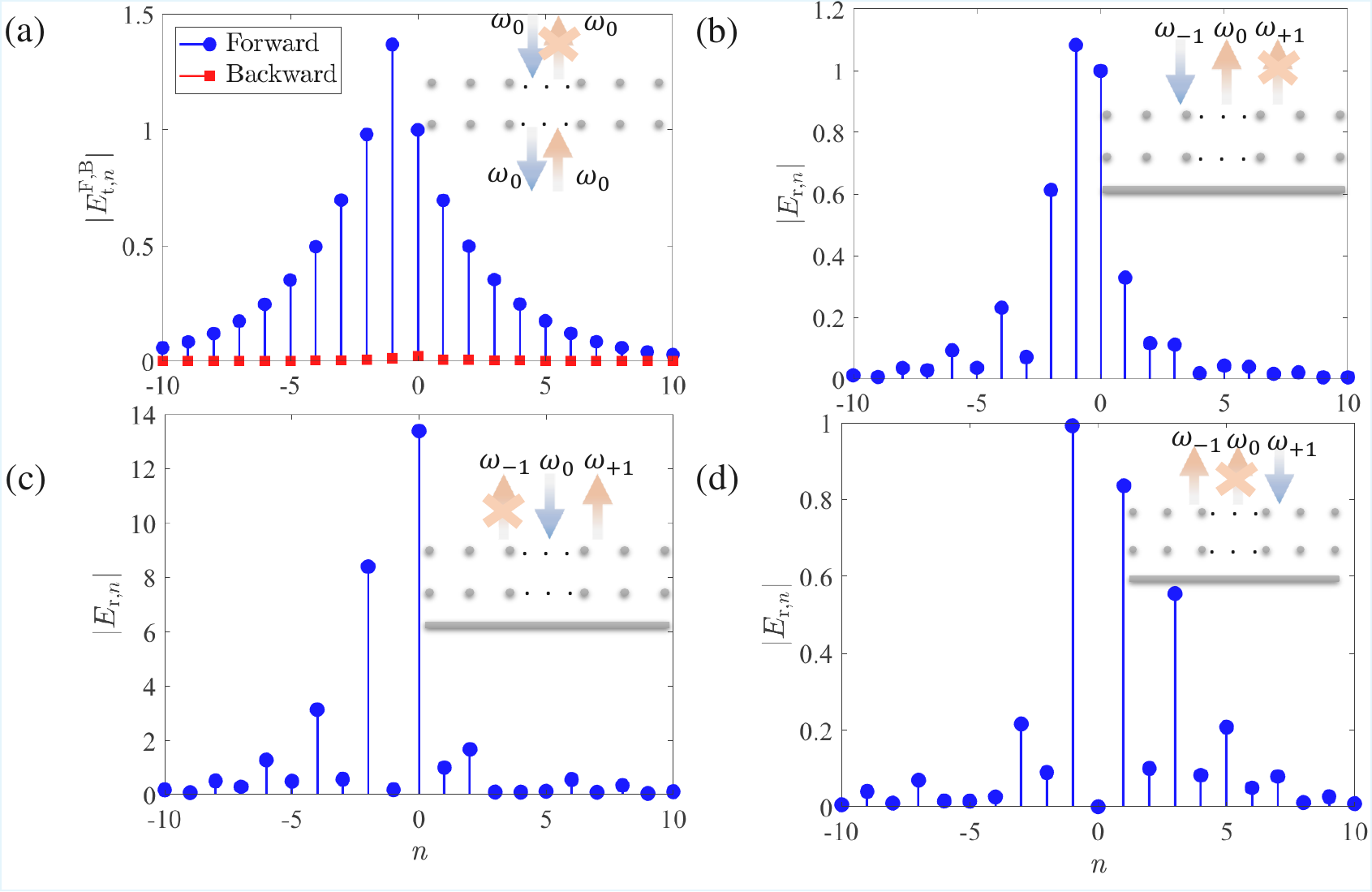}
  \caption{(a) Distribution of complex amplitudes of different transmitted harmonics for the optimized metasurface stack providing isolation at the fundamental frequency.  The inset depicts the desired functionality of the cascade consisting of two time-varying metasurfaces. Blue and red stems  
  correspond to the forward and backward illumination scenarios, respectively. (b)--(d) Distribution of complex amplitudes of different reflected harmonics for the optimized metasurface stack providing temporal circulation. Blue and light red arrows in the insets denote the incident and optimized reflected frequency harmonics, respectively.}
  \label{fig:results}
\end{figure}

The amplitudes of different transmitted harmonics for the optimized metasurface stack are shown in~Fig.~\ref{fig:results}(a). The transmitted harmonics for the forward illumination are shown with blue stems. We can observe a spectrum with a Gaussian-type frequency distribution where the amplitude of the transmitted $\omega_0$ harmonic is very close to 1. On the contrary, for the backward illumination (red stems), the transmission is strongly suppressed at all frequencies, with the transmitted electric field at $\omega_0$ being equal to $0.021$~V/m. The value of the cost function represented in~Eq.~\eqref{eq:error_metric} after optimization is $0.0211$.
Thus, the isolation ratio at the fundamental frequency reaches $33.56~\rm{dB}$. It is worth noting that, depending on the application, sideband frequency harmonics ($\omega_n$ with $n \neq 0$) possessing non-zero amplitudes in the forward illumination scenario may be undesirable. In this case, they can be easily suppressed by incorporating a conventional frequency band-pass filter into the metasurface stack.

\subsection{Temporal circulation}

To further demonstrate the versatility of the proposed stacked metasurfaces, we design a novel and exotic system. In microwave and optical engineering, one of the fundamental nonreciprocal components is the circulator—a device that operates at a single frequency but routes signals nonreciprocally among three distinct spatial ports in a unidirectional, cyclic manner~\cite{pozar2021microwave}. In this section, we propose a temporal analogue of the circulator. Unlike conventional circulators, this device features a single spatial port but three distinct frequency ports, enabling the wave to circulate nonreciprocally across these frequencies in a manner analogous to spatial circulators. We term this device a \textit{temporal (or frequency) circulator}. Potential applications include compact nonreciprocal light guiding, the realization of synthetic dimensions~\cite{yuan2018synthetic,buddhiraju2021arbitrary}, and dynamic spectral multiplexing for integrated photonic systems.

To ensure that there is only one spatial port, we incorporate a perfectly conducting plate (ground plane) behind the metasurface stack, as shown in the inset of Fig.~\ref{fig:results}(b). This way, we truncate half of the space below the plate and consider the normally incident and reflected waves to share the same single spatial port. 
For the three frequency (temporal) ports, we consider three frequencies coupled through the temporal modulations in the metasurfaces: $\omega_0$ and $\omega_{\pm 1} = \omega_0 \pm \omega_{\rm m}$. 
The distance between the conducting plate and the closer metasurface is denoted as $D_2$. 
In the transfer-matrix method, we model the conducting plate as a parallel short circuit with the ABCD matrix $\overline{\overline{T}}_{\rm pl} $ being equal to $\overline{\overline{T}}_{\rm m}$ in Eq.~(\ref{eq:T_matrix_definition}) where $\overline{\overline{Y}}_{\mathrm{g}}$ is replaced by a diagonal matrix $\alpha \overline{\overline{I}} $ (here $\alpha \to \infty$).

For the operation of the temporal circulator, we define cost functions based on the desired reflection behavior across the three temporal harmonics. Each incident wave is assumed to have a $1~\text{V/m}$ electric-field amplitude. When the metasurface stack is excited by one harmonic (e.g., $\omega_{-1}$), the reflected wave should appear entirely at one of the other two considered harmonics (say, $\omega_{0}$) with the same $1~\text{V/m}$ amplitude, while the remaining harmonic ($\omega_{+1}$) is completely suppressed. This property ensures that waves at each input frequency are uniquely routed to another frequency port without amplitude loss. By cyclically permuting the input among the three harmonics, the same nonreciprocal mapping must hold. The cost functions that penalize deviations from this ideal circulation (unit-amplitude transfer to the designated port and zero leakage to the undesired port) are chosen as
\begin{subequations}
\label{eq:z_series}
\begin{align}
\Delta_{1} &= \left| \, \left| {E_{\rm r,0}^{\rm F}} \right| - 1 \, \right|
      + \left| \, \left| {E_{\rm r,+1}^{\rm F}}  \right| - 0 \, \right|
      \quad \text{(excitation at $\omega_{-1}$)}
\label{eq:z1_formula} \\[2mm]
\Delta_{2} &= \left| \, \left| {E_{\rm r,+1}^{\rm F}}  \right| - 1 \, \right|
      + \left| \, \left| {E_{\rm r,-1}^{\rm F}}  \right| - 0 \, \right|
      \quad \text{(excitation at $\omega_{0}$)}
\label{eq:z2_formula} \\[2mm]
\Delta_{3} &= \left| \, \left| {E_{\rm r,-1}^{\rm F}}  \right| - 1 \, \right|
      + \left| \, \left| {E_{\rm r,0}^{\rm F}}  \right| - 0 \, \right|
      \quad \text{(excitation at $\omega_{+1}$)}
\label{eq:z3_formula}
\end{align}
\end{subequations}
The total error function of the circulation optimization problem is defined as a summation of all these three functions, that is, $\Delta_{\rm circ}=\Delta_1+\Delta_2+\Delta_3$. The periodicity $d$ and radii $r_0$ are chosen to be the same as in the previous example.

Figures~\ref{fig:results}(b)--(d) show the reflection coefficients of waves at different frequency harmonics for the optimized metasurface stack for three different illumination scenarios (at $\omega_0$ and $\omega_{\pm 1}$).  
The optimized material parameters are listed in the Supplementary Material.
Blue arrows in insets of the plots show the excitation frequency, while the light red arrows stand for the optimized reflected frequency harmonics. As we can see from Fig.~\ref{fig:results}(b), the incident $\omega_{-1}$ harmonic is reflected towards the $\omega_0$ temporal port with 1~V/m field amplitude, while the reflected $\omega_{+1}$ harmonic is suppressed to 0.3291~V/m. 
On the other hand, incident $\omega_{0}$ harmonic is reflected towards the $\omega_{+1}$ temporal port with 0.9991~V/m field amplitude, as is seen in Fig.~\ref{fig:results}(c) (note the large scale of the vertical axis in the figure). The reflected $\omega_{-1}$ harmonic reaches 0.1895 V/m. 
Finally, incident harmonic $\omega_{+1}$ is reflected predominantly to the $\omega_{-1}$ temporal port (0.9919~V/m) and minimally coupled to the $\omega_{0}$ port (0.001~V/m), as shown in Fig.~\ref{fig:results}(d).
The value of the final cost function introduced as the sum of the error functions in~Eq.~\eqref{eq:z_series} after optimization is $0.1739$.
The isolation levels for the three incident scenarios mentioned above are $-9.90 \rm{dB}$, $-14.90 \rm{dB}$, and $-60 \rm{dB}$, listed in the same order. It should be mentioned that the optimized metasurface stack is not matched and exhibits also non-zero ``retro'' reflections, that is, reflections to the same temporal port from which the incident wave arrived. These could be minimized by incorporating additional conditions in the cost function and increasing the number of metasurfaces in the stack. Non-zero reflected sideband harmonics beyond $\omega_0$ and $\omega_{\pm 1}$ can be suppressed in practical settings with bandpass frequency filters.


\section{Proposed solution for waveguide realization}

\begin{figure}\includegraphics[width=0.95\linewidth]{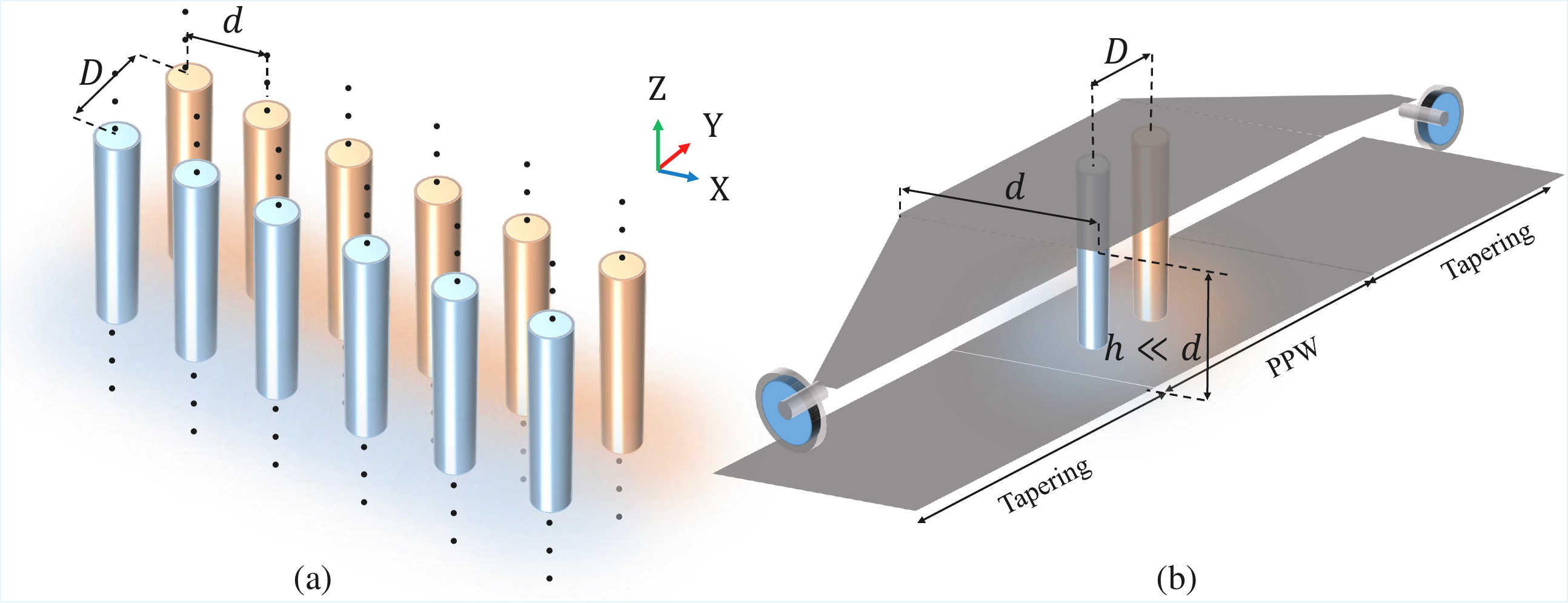}
  \caption{(a) Two time-varying metasurfaces in free space and (b) compact PPW configuration emulating these metasurfaces with only two sub-wavelength meta-atoms. Two extra sections are added at both sides of the PPW in order to effectively convert the TEM dominant mode of PPW and coaxial cables.}
  \label{fig:realization}
\end{figure}

In this section, we introduce an efficient and straightforward solution for the microwave realization of the proposed stacked metasurfaces. Without loss of generality, in our illustration, we consider a stack consisting of two metasurfaces, as shown in Fig.~\ref{fig:realization}(a). However, the method is generic and applies to an arbitrary number of metasurfaces. A linearly-polarized incident wave, propagating in the normal direction to the metasurfaces plane (with the wave vector along the $y$-axis), has its electric field component aligned parallel to the $z$-axis, while the magnetic field is perpendicular to the cylinders ($x$-axis).
Naturally, the realization of the metasurface shown in Fig.~\ref{fig:realization}(a) is very cumbersome as it requires many time-modulated cylindrical meta-atoms with lengths much larger than the operational wavelength. 

To implement such structures in practice, one can consider an equivalent system inside a parallel-plate waveguide (PPW) and exploit the image theory principle~\cite{cheng1989field}. Let us consider a PPW configuration shown in Fig.~\ref{fig:realization}(b). Here, to represent each metasurface, we need only one cylindrical meta-atom with the same radius $r_0$ and a finite (sub-wavelength) height $h$ being placed between two conducting plates. 
The width of these plates in the $x$-direction must match the periodicity of the array, that is, $d$. The central region of the structure functions as a PPW supporting the TEM mode. Importantly, in the absence of side walls, the magnetic field at the PPW opening is oriented along the $x$-direction (with vanishing $y$-component), if the $h \ll d \ll \lambda_0$~\cite{saarela2002ams}. 
Therefore, such a PPW configuration effectively emulates perfect electric and magnetic conductor walls separated by $h$ in the $z$-direction and by $d$ in the $x$-direction, respectively. This feature allows us to emulate the infinite geometry in Fig.~\ref{fig:realization}(a) by a finite geometry shown in Fig.~\ref{fig:realization}(b). The cascade of two metasurfaces is modeled by a cascade of two meta-atoms separated by a distance $D$. The height of the PPW can be chosen such that the waveguide characteristic impedance becomes matched to that of standard coaxial connectors, that is, $50\,\Omega$. For ensuring perfect mode conversion between the PPW and the coaxial connector (supporting the dominant cylindrical TEM mode), a tapering section can be designed, as shown in Fig.~\ref{fig:realization}(b). The height and width of the tapering opening section should be configured to maintain constant impedance throughout the entire transition region. The equivalence between the scattering parameters of the free-space system in Fig.~\ref{fig:realization}(a) and the PPW system in Fig.~\ref{fig:realization}(b) is demonstrated in the Supplementary Material.

\section{Conclusion}

In this paper, a general theoretical framework for designing stacked time-varying metasurfaces is presented, along with several demonstrators of nonreciprocal devices. The plasma and collision frequencies are selected as the modulated material parameters, with material dispersion taken into account. Accurate circuit models are proposed to interpret the behavior of both individual cylindrical meta-atoms and their arrays. Establishing the equivalent circuit model enables straightforward computation of the transfer matrices for time-varying metasurfaces within a cascade.  Finally, a practical realization method is introduced using a parallel-plate waveguide configuration. The proposed stacked metasurfaces offer a compact, versatile, and scalable route toward spatiotemporal platforms for next-generation photonic and microwave systems. The developed methodology is not limited to arrays of cylinders and can be used to cascades of any metasurfaces modeled by sheet impedances that are periodically modulated in time. 

Further generalization of this approach can be achieved by incorporating spatial non-uniformity in the in-plane metasurface direction. In this respect, it is important to stress the fundamental advantages of the proposed time-modulated nonreciprocal metasurface structures compared to earlier developed solutions based on fully reflective metasurfaces (for example, as in \cite{wang2020theory}). It is well established that in order to fully control anomalous reflections into desired directions, it is necessary to carefully engineer distributions of high-order evanescent Floquet harmonics \cite{asadchy2016perfect, diaz2017generalized}. This requirement, applicable also to nonreciprocal wave transformations, demands the use of dense subwavelength arrays of different time-modulated meta-atoms, which are difficult to optimize and realize. The fundamental reason for these difficulties is that the optimal anomalous reflectors must exhibit a non-local response. In contrast, full control of reflection-free transmission to any angle is possible to achieve using locally-responding arrays~\cite{asadchy2016perfect}. This means that the developed design method is also directly applicable to spatially non-uniform time-modulated arrays for transmission control. Since it is enough to optimize only the local response, each meta-atom can be optimized separately in an infinite array of identical meta-atoms, as is done in this paper for uniform arrays. Thus, we expect that this method will open a way to realize nonreciprocal metasurfaces that will be able to control also the direction of propagation of transmitted waves.




\medskip
\textbf{Acknowledgements} \par 
MM, ST, and VA acknowledge the Horizon 2020 Framework Programme (PULSE). ST and VA acknowledge the Research Council of Finland within the RCF-DoD Future Information Architecture for IoT initiative under grant no. 365679. X.W. acknowledges the Fundamental Research Funds for the Central Universities, China (project no. 3072024WD2603)

\medskip

%
\bibliographystyle{MSP}
\bibliography{nonreciprocity_refs}

\end{document}